\documentclass{article}
\usepackage{spconf,amsmath,graphicx,hyperref}
\usepackage{multirow}
\usepackage[table,xcdraw]{xcolor}
\usepackage[normalem]{ulem}
\useunder{\uline}{\ul}{}
\usepackage{amssymb}
\usepackage{graphicx} 
\usepackage{caption}  
\usepackage{comment}

\title{INDEX-MSR: A HIGH-EFFICIENCY MULTIMODAL FUSION FRAMEWORK FOR SPEECH RECOGNITION}
\name{Jinming Chen, Lu Wang, Zheshu Song, Wei Deng}
\address{Artificial Intelligence Platform Department, bilibili, China \\
\{chenjinming, wanglu08, songzheshu, xuanwu\}@bilibili.com}
\begin{document}
%
\maketitle
\begin{abstract}

Driven by large-scale datasets and LLM-based architectures, automatic speech recognition (ASR) systems have achieved remarkable improvements in accuracy. However, challenges persist for domain-specific terminology, and short utterances lacking semantic coherence, where recognition performance often degrades significantly.
In this work, we present Index-MSR, an efficient multimodal speech recognition framework. At its core is a novel Multimodal Fusion Decoder (MFD), which effectively incorporates text-related information from videos (e.g., subtitles and presentation slides) into the speech recognition. This cross-modal integration not only enhances overall ASR accuracy but also yields substantial reductions in substitution errors.
Extensive evaluations on both an in-house subtitle dataset and a public AVSR dataset demonstrate that Index-MSR achieves state-of-the-art accuracy, with substitution errors reduced by 20–50\%.
These results demonstrate that our approach efficiently exploits text-related cues from video to improve speech recognition accuracy, showing strong potential in applications requiring strict audio-text synchronization, such as audio translation.

\end{abstract}
\section{Introduction}
\label{sec:intro}

In recent years, the availability of large-scale speech datasets and the introduction of LLM-based recognition frameworks into ASR models have led to significant improvements in accuracy \cite{Mu_Shao_Wei_Yu_Xie, Prabhavalkar_Hori_Sainath_Schl}. Large-scale multilingual pre-trained models \cite{kashiwagi2025whale} have helped narrow the performance gap between high-resource and low-resource languages. However, for accented speech, unclear pronunciation, named entities, and short utterances with limited semantic coherence, the accuracy of speech-only recognition still degrades considerably.

In ASR applications, an important scenario involves the word-level transcription of speech embedded in videos. Videos frequently contain abundant text-related cues (e.g., subtitles and presentation slides) related to the spoken content, which can substantially complement the speech signal and are available at scale across multiple languages.  Leveraging visual information in these video scenarios thus holds significant potential for enhancing the robustness and generalization of ASR models.

However, directly aligning text information from visual content with ASR transcripts for correction remains challenging. Even the most relevant subtitles often undergo rephrasings (e.g., shortening, paraphrasing), stylistic edits, and temporal misalignments, posing challenges for achieving precise word-level recognition \cite{Poncelet_Van}. Moreover, videos typically include substantial non-subtitle textual content, making subtitle extraction inherently challenging. Errors arising from this process can significantly compromise the alignment quality in ASR. Importantly, visual cues are not limited to subtitles; for example, domain-specific terminology appearing in slides or presentations can also provide valuable references to improve ASR accuracy. Consequently, a generalizable multimodal speech recognition approach holds significant importance.

Some multimodal large language models (MLLMs) \cite{xu2025qwen2,peng2024survey} can incorporate both audio and video information to provide ASR results. However, when subtitles and audio are misaligned, such methods tend to favor subtitle content over accurate word-level transcription, limiting their applicability in scenarios requiring precise temporal alignment. Moreover, they typically demand substantial computational resources, as they rely on large-scale language models (LLMs) to interpret multimodal inputs.
Therefore, efficiently leveraging video information to enhance ASR accuracy while maintaining its intrinsic temporal alignment is highly valuable.

In this work, we propose Index-MSR, an innovative and efficient multimodal alignment model that preserves the strict temporal alignment capability of ASR while effectively leveraging visual information to significantly enhance recognition performance.
The main contributions of this work are:

1. We investigate efficient, low-resource, end-to-end multimodal ASR training methods that integrate text-related features from video into verbatim speech recognition.

2.We propose a novel Multimodal Fusion Decoder (MFD) that substantially enhances ASR accuracy and achieves a significant reduction in substitution errors in the multimodal recognition scenario.

3. We conduct extensive evaluations on both in-house and open-source datasets, comparing against state-of-the-art models and MLLM methods.

4. We identify promising potential applications of the proposed technique in source-audio translation scenarios.

\section{Methods}
\label{sec:format}

Our work is built upon an end-to-end (E2E) ASR system \cite{Dong_Xu_Xu_2018}, where the textual information is derived from weakly supervised OCR features. The feature extraction encoder is kept frozen during training and does not perform any explicit discrimination between subtitle text and background text. To preserve the capability for verbatim transcription, we adopt a Conformer-based E2E ASR architecture. The key innovation lies in our Multimodal Fusion decoder (MFD) design, which enables efficient fusion of multimodal information and leads to substantial improvements in recognition accuracy.

\subsection{End-to-end ASR Enocoder}
\label{ssec:subhead}

\begin{figure*}[htbp]  
	\centering  
	\includegraphics[width=0.85\textwidth]{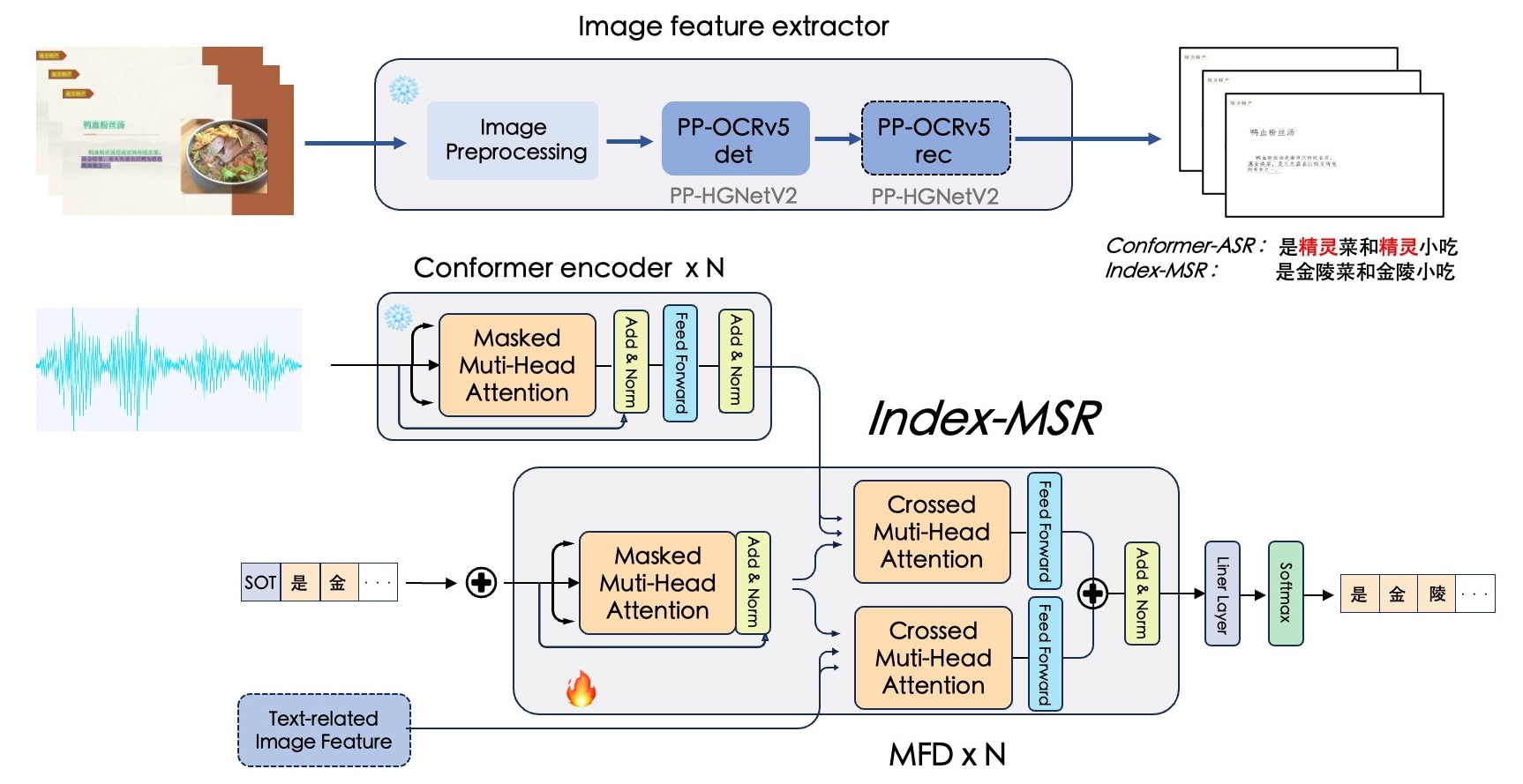}  
	\caption{Index-MSR: Architecture and Processing Pipeline}  
	\label{fig1}  
\end{figure*}

We adopt a Conformer-based encoder \cite{gulati2020conformer} shown in Fig.\ref{fig1}, which integrates global self-attention with local convolution operations to capture both long-range dependencies and fine-grained local patterns in the speech signal. To regularize the encoder and guide it towards acoustically meaningful outputs, we apply a Connectionist Temporal Classification (CTC) objective on both its encoder layer outputs. The CTC objective enforces a monotonic alignment between the encoder features.

\subsection{OCR-Integrated Text Feature Encoder}
\label{ssec:subhead}

We extract visual textual features from video frames using a SOTA OCR pipeline based on PP-OCRv5 \cite{cui2025paddleocr30technicalreport}. First, an image preprocessing module enhances input image quality and corrects distortions or misorientations. Next, the text detection model using an advanced PP-HGNetV2 backbone. Detected text lines are then passed through a text line orientation classification model. Finally, the text recognition model encodes each text line into feature representations. A lightweight CTC-based branch for efficient decoding of text features. The image features extracted prior to CTC alignment are used for feature representation, offering a compact and semantically enriched embedding well-suited for multimodal fusion in ASR.
Importantly, we did not perform any dataset-specific optimization on the OCR pipeline, nor did we distinguish whether the extracted textual information originated from subtitles or background content; instead, all visual features were uniformly packed and fed into the MFD module.

\subsection{Multimodal Fusion Decoder (MFD)}
\label{ssec:subhead}

Although both audio and visual encoders adopt CTC alignment, their features come from different domains, temporal for speech and spatial for images. Our goal is strict time-aligned transcription; thus, we design a spatiotemporal interleaved cross-attention decoder that fuses features across domains while preserving word-level alignment.
In our proposed Multimodal Fusion Decoder (MFD), we extend the cross-attention mechanism within a transformer-based multimodal decoder architecture. Specifically, audio embeddings from the speech encoder $(T)$ and visual embeddings $(I)$ from the OCR-based text encoder are jointly injected into the decoder’s attention module. By integrating these heterogeneous modalities together with historical target information, MFD produces context representations through the cross-modal fusion mechanism, followed by the feed-forward layer and softmax prediction. These representations jointly encode information from both modalities and are then used to construct the output decoding sequence. This design not only preserves the strict temporal alignment inherent in ASR but also incorporates domain-specific textual information from visual content, thereby significantly improving recognition accuracy—particularly for named entities and rare words. (Fig.\ref{fig1})

\begin{align}
	Q_t &= \mathbf{Q} W_{q}^{t}, \quad 
	K_t^{a} = \mathbf{T} W_{k}^{t}, \quad 
	V_t^{a} = \mathbf{T} W_{v}^{t} \\[6pt]
	\mathrm{head}_t &= \mathrm{Attention}(Q_t, K_t^{a}, V_t^{a}) 
	= \mathrm{softmax}\!\left(\frac{Q_t K_t^{a}}{\sqrt{d_k}}\right) V_t^{a} \\[8pt]
	Q_t &= \mathbf{Q} W_{q}^{i}, \quad 
	K_i ^{v}= \mathbf{I} W_{k}^{i}, \quad 
	V_i^{v} = \mathbf{I} W_{v}^{i} \\[6pt]
	\mathrm{head}_i &= \mathrm{Attention}(Q_t, K_i^{v}, V_i^{v}) 
	= \mathrm{softmax}\!\left(\frac{Q_t K_i^{v}}{\sqrt{d_k}}\right) V_i^{v} \\[8pt]
	\mathbf{H} &= \mathrm{head}_t + \mathrm{head}_i  \tag{5}
\end{align}

For the multimodal fusion mechanism, we adopt a dual-modality injection strategy, where both audio features $(T)$  and video features $(I)$  are incorporated via cross-attention with the labeled textual representations $(Q)$  during training.  
\section{EXPERIMENTS}
\label{sec:pagestyle}

\subsection{Datasets}
We conduct extensive experiments on both an in-house dataset and the publicly available Chinese-Lips dataset \cite{Zhao_Jia_Wang_Zhou_Wang_Qin} to evaluate the effectiveness of our proposed approach.

The in-house dataset consists of videos with embedded subtitles and word-level aligned speech annotations. In total, it contains approximately 190 hours of data, of which 171 hours are used for training and 19 hours for validation. The test set, comprising 8.6 hours of subtitled video with word-level aligned speech annotations, is isolated from the training data and used to evaluate ASR accuracy.

Chinese-Lips dataset \cite{Zhao_Jia_Wang_Zhou_Wang_Qin}. This is a large-scale multimodal AVSR dataset comprising around 100 hours of speech, video, and corresponding manual transcriptions. The visual modality in this dataset covers both lip-reading information and presentation slides used by the speakers, as well as background scenes and other visual contexts. 

\subsection{Experiment configuration}
For the visual modality, we directly employ the recognition backbone of PP-OCRv5 \cite{cui2025paddleocr30technicalreport} as the text feature extractor. Importantly, the OCR encoder is used without any additional fine-tuning on our datasets. All text-related information is packed at the sentence level and then used as the input for the text modality.

Our speech recognition backbone is an attention-based encoder–decoder (AED) architecture implemented with the WeNet toolkit \cite{Yao_Wu_Wang_Zhang_Yu_Yang_Peng_Chen_Xie_Lei_2021}. The encoder adopts a Conformer structure with 12 blocks (a1: 8 attention heads, 2048 linear units; a2: 16 attention heads, 4096 linear units), while the decoder is a Transformer-based module with 3 blocks (a1: 8 attention heads, 2048 linear units; a2: 16 attention heads, 4096 linear units). The model is initialized from a pre-trained checkpoint trained on the Mandarin–English bilingual speech data, providing strong acoustic representations prior to multimodal fine-tuning. During multimodal alignment, the speech encoder parameters are frozen to ensure that performance improvements stem from the proposed multimodal fusion rather than additional training of the acoustic backbone.

\begin{table*}[t] 
	\centering
	\caption{ASR performance of different modality models on the in-house subtitle dataset.}
	\renewcommand{\arraystretch}{1.2} 
	\setlength{\tabcolsep}{6pt}       
	\begin{tabular}{c|c|c|c|cccc}
\hline
&                                         &                        &                            & \multicolumn{4}{c}{WER}                                        \\
\multirow{-2}{*}{ID} & \multirow{-2}{*}{Model}                 & \multirow{-2}{*}{Size} & \multirow{-2}{*}{Modality} & Substitution  & Deletion     & Insertion    & Overall          \\ \hline
a1                   & Conformer-ASR-S                         & 126M                   & Speech                     & 5650          & 664          & 506          & 10.36 \%         \\
m1                   & Conformer-ASR+OCR                       & 126M                   & Speech+Visual              & 4364          & 553          & 836          & 8.73 \%          \\
m2                   & Index-MSR-S                             & 140M                   & Speech+Visual                     & 2772          & 556          & 835          & 6.32 \%          \\
 \hline
a2                   & Conformer-ASR-L                         & 432M                   & Speech              & 3740          & 842          & {\ul 382}          & 7.54 \%          \\
m3                   & Index-MSR-L                             & 464M                   & Speech+Visual              & \textbf{1850} & \textbf{406} & \textbf{619} & \textbf{4.37 \%} \\ \hline
a3                   & Paraformer-Large                        & 220M                   & Speech                     & 3746          & 1315         & 585          & 8.58 \%          \\
a4                   & {\color[HTML]{000000} FireRedASR-AED-L} & 1.1 B                  & Speech                     & {2230}          & {\ul 339}          & 625          & 4.85 \%          \\
m4                   & Gemini-2.5-Pro                          & 175B                   & Speech+Visual              & {\ul 1762}          & 1135         & 12518        & 23.42\%          \\ \hline
	\end{tabular}
	\label{tab1}
\end{table*}

\subsection{Analysis and Ablation}

Our proposed MFD structure is trained on the in-house audio-labeled subtitle dataset. To ensure that performance gains are not simply due to additional training data, we also trained a Conformer-based ASR baseline using only the speech modality from the same dataset (a1, a2 in Table 1). These serve as direct unimodal comparisons for validating the benefit of multimodal fusion.

For Index-MSR, we used the same training data as a1/a2 but additionally incorporated textual features from video, with results reported as m2 and m3. To further probe alignment effectiveness, we conducted a control experiment where raw OCR outputs from PP-OCRv5 were directly embedded as textual features (m1). While m1 improves WER over a1, its substitution error reduction is far smaller than m2, underscoring the importance of high-quality textual feature extraction for multimodal alignment.

In contrast, Index-MSR with MFD consistently outperforms unimodal baselines under the same training conditions, achieving nearly 50\% fewer substitution errors. Error analysis shows that most corrected substitutions involve named entities and homophones—precisely the categories our fusion design targets.

To examine scalability, we tested larger ASR encoders initialized from a 30,000 hour bilingual corpus. While a stronger encoder improves unimodal accuracy (a2), MFD still delivers ~50\% substitution error reduction (m3), demonstrating robustness across encoder capacities. Moreover, with only ~200 hours of multimodal data, our approach matches or surpasses state-of-the-art speech-only models trained on tens of thousands of hours (e.g., a 1.1B-parameter AED system, a4 in Table 1) \cite{Xu_Xie_Tang_Hu}.

Finally, we compared against large multimodal models. In m4, Gemini 2.5 Pro jointly processes video and audio, achieving the lowest substitution errors by leveraging subtitle text. However, in cases of video–audio mismatch, it often inserts subtitle words absent from speech, leading to high insertion errors.

\subsection{Generalization Experiments}
In addition to validating the generalization and scalability of our model on the in-house dataset, we further evaluated it on the Chinese-Lip dataset, which consists of multimodal video with weakly supervised textual signals derived from presentation slides.

The Chinese-LiPS dataset provides benchmark baselines \cite{Zhao_Jia_Wang_Zhou_Wang_Qin} for both speech only and multimodal speech recognition. For the speech only setting, the Whisper large-v2 \cite{radford2023robust} model is adopted, achieving a CER of 3.99\% when trained and evaluated solely on audio signals. For multimodal baselines, Whisper is extended with additional inputs derived from the video stream: text extracted from presentation slides using PP-OCR, and semantic keywords obtained with InternVL2  \cite{chen2024expanding,gao2024mini, chen2024far, chen2024internvl}. These multimodal variants serve as the official baselines for evaluating ASR performance on Chinese-LiPS.

We compared models of  the Index-MSR-L against both unimodal and multimodal baselines. Consistent with our in-house results, the proposed MFD significantly improves alignment under weak supervision. Specifically, we observed a 20\% relative reduction in substitution errors, which are especially prevalent in named entities and domain-specific terminology. Compared to baseline multimodal approaches that simply concatenate speech and image features, our MFD achieves both the lowest substitution error rate and the lowest overall WER.  (Table 2)

\section{Conclusion}
\label{sec:typestyle}

We propose Index-MSR, an efficient multimodal fusion framework that leverages limited alignment data to calibrate ASR outputs with visual cues, reducing substitution errors by 50\% and effectively handling named entities and rare words. We further validate the MFD by scaling ASR models and training on weakly supervised multimodal datasets, where it consistently improves accuracy, confirming its robustness and scalability.
Compared with SOTA unimodal ASR and multimodal large models, Index-MSR uniquely combines low-resource efficiency, high accuracy, and fine-grained word-level alignment with timestamps, making it well-suited for verbatim ASR in video-rich environments. It is particularly promising for source-audio translation, where precise timing and accurate entity recognition are critical.

\begin{table}[h]  
	\centering      
	\caption{ASR performance of different modality models on the Chinese-LiPS dataset \cite{Zhao_Jia_Wang_Zhou_Wang_Qin}.}
	\renewcommand{\arraystretch}{1.2} 
	\setlength{\tabcolsep}{2.5pt}       
\begin{tabular}{cccccl}
	\hline
	\textbf{ID}               & \textbf{Modality}                      & \textbf{S}                  & \textbf{D}                 & \textbf{I}                 & WER                \\ \hline
	b1                        & Speech only                            & 3851                        & 1697                       & 437                        & 3.99 \%            \\
	b2                        & Speech + Slides                        & 3531                        & 447                        & 510                        & 2.99 \%            \\
	b3                        & Speech + Lip                           & 4499                        & 509                        & 522                        & 3.69 \%            \\
	b4                        & Speech + Lip + Slides                  & 3047                        & 335                        & 484                        & 2.57 \%            \\ \hline
	{\color[HTML]{656565} a2} & {\color[HTML]{656565} Conformer-ASR-L} & {\color[HTML]{656565} 3088} & {\color[HTML]{656565} 514} & {\color[HTML]{656565} 599} & 2.80 \%            \\
	\textbf{m3}               & \textbf{Index-MSR-L}                   & \textbf{2483 ↓}             & \textbf{233 ↓}             & \textbf{413 ↓}             & \textbf{2.08 \% ↓} \\ \hline
\end{tabular}
\end{table}

\bibliographystyle{IEEEbib}
\bibliography{strings,refs}

\end{document}